# The Greater The Power, The More Dangerous The Abuse: Facing Malicious Insiders in The Cloud

Nikolaos Pitropakis
School of Electrical and Computer Engineering
Georgia Institute of Technology
Atlanta, United States of America
e-mail: pitropakis@gatech.edu

Christos Lyvas, Costas Lambrinoudakis
Department of Digital Systems
University of Piraeus
Piraeus, Greece
e-mail: {clyvas,clam}@unipi.gr

*Abstract*—The financial crisis made companies around the world search for cheaper and more efficient solutions to cover their needs in terms of computational power and storage. Their quest came to end with the birth of Cloud Computing infrastructures. However, along with the new promising technology, new attack vectors were born, and one old and known threat, that of Malicious Insiders reappeared. Insiders can use their privileged position inside the Cloud infrastructure to accomplish or help in attacks against a Cloud infrastructure. In this paper, we propose a practical and efficient intrusion detection system solution for Cloud infrastructures based on Graphical Processing Unit (GPU) acceleration. Our solution monitors the deployed virtual machines operations and especially those of the host Operating System's, known as Dom0, correlating the collected information to detect uncommon behavior based on the Smith-Waterman algorithm. Our proposal makes possible the cooperation of a variety of known hypervisors along with every known GPU acceleration unit used, thus offering the maximum of security mechanics while at the same time minimizing the imposed overhead in terms of Central Processing Unit (CPU) usage.

*Keywords-Cloud Computing; Security; Malicious Insider; IDS; GPU Acceleration.*

## I. INTRODUCTION

While economic growth is considered low in the vast majority of the global market, Cloud Computing infrastructures have grown beyond imagination. Their revenues jumped by 25% for 2016 with strong estimated growth ahead. Leading Amazon Web Services (AWS) and Microsoft Azure grew 53% in 2016 [9]. AWS, which introduced the concept of Cloud Computing managed to generate revenue of 13 billion dollars in 2016. As migration services become more convenient and at the same time more appealing, more companies will choose the pay-per-use model that Cloud Computing offers.

Cloud Computing by design cannot offer physical isolation among Virtual Machines (VMs), since all resources are shared. Various attack vectors have been developed [24] and continue to be updated following the lead of security experts, trying to identify shared resources and gain unauthorized access to them. Hypercall attack injection [18], co-residency detection, shared memory vulnerabilities [26] and privilege escalation [7], are only a few examples of the attack vectors that could harm the confidentiality, integrity and availability of Cloud systems and data. It is a fact that Cloud infrastructure's attack surface is an expanded version of older Information Technology (IT) infrastructures, because a potential adversary can make use of additional attacking points to explore a vulnerability (e.g., a VM, a management platform or other components). Malicious Insider threat has reappeared and has become the main reason for data leakage as 1 out of 3 organizations have experienced an insider attack in the year 2016 [10].

Several approaches have been proposed to augment security in Cloud infrastructures. Most of them inherit their operational methodologies from conventional IT systems. The most popular approaches among the community try either to scatter the information among the whole infrastructure (in terms of data storage) [13] or implement multiple Intrusion Detection Systems (IDS) [17] and audit mechanisms [15]. Several of them monitor system calls to detect malicious activities [2][25][29]. The recent trend is to migrate the entire VM to another part of the infrastructure, thus forcing the potential attacker to be one step behind [43]. Most of them are unable to detect attacks against the Cloud from privileged users and especially attacks, which are orchestrated by multiple VMs.

Thus, we introduce Modified And Deterring Realtime Observation Wards (MAD CROW) for detecting malicious activities against the VM and against the Cloud infrastructure itself. The principle of our approach is to monitor the hypercalls of the VMs independently and the system calls of the privileged domain (Dom0 in XEN [41], Virtual Machine Manager (VMM) in Kernel-based Virtual Machine (KVM) [14]), in a way similar to a host based IDS, combining all gathered information to protect each VM and the whole Cloud infrastructure at the end of the day.

To be more specific, we make use of mechanisms that trace hypercalls (Xentrace in the case XEN [42], Perfm KVM in the case of KVM [20]) and systemcalls (strace command [34]) and process them in order to generate attack patterns and process abnormal behaviors. In contrast to other cloud IDSs [5] that use machine learning classifiers as black-box, the proposed system generates attack patterns using the Smith-Waterman algorithm [30] and performs similarity tests between the attack patterns and the data (hypercalls and system calls) collected to decide whether the cloud infrastructure is under attack or not, with a certain level of confidence. Since our approach operates on the Cloud infrastructure as a service layer, in a transparent manner, no modifications to the underlying layers are required.

Overall, the contributions of the paper could be summarized as follows:
- We introduce a hybrid solution, which depends on hypercalls and system calls to detect abnormal behavior in Cloud infrastructures.
- We enhance the performance of this solution using GPU acceleration instead of CPU computational resources
- Our solution is adaptable depending on the resources (GPU) and the Cloud infrastructure (hypervisor used)

The rest of the paper is organized as follows: Section II offers some background information while Section III provides a related literature review. Section IV introduces the malicious insider threat model. Section V presents our approach to detect malicious activities in Cloud infrastructure. Finally, Section VI draws the conclusions giving some pointers for future work as well.

## II. BACKGROUND

### A. Hypervisors

A hypervisor is in most cases a software, which acts as a layer between the hardware and the VMs. Basically, it is a level of abstraction that isolates either operating systems or applications from the underlying computer hardware. This abstraction allows the underlying host machine hardware to independently operate one or more virtual machines as guests, allowing multiple guest VMs to effectively share the system's computational resources, such as processor, memory, storage, network bandwidth, etc. There are two implementations of the hypervisor concept worth mentioning, one is XEN and the other is KVM.

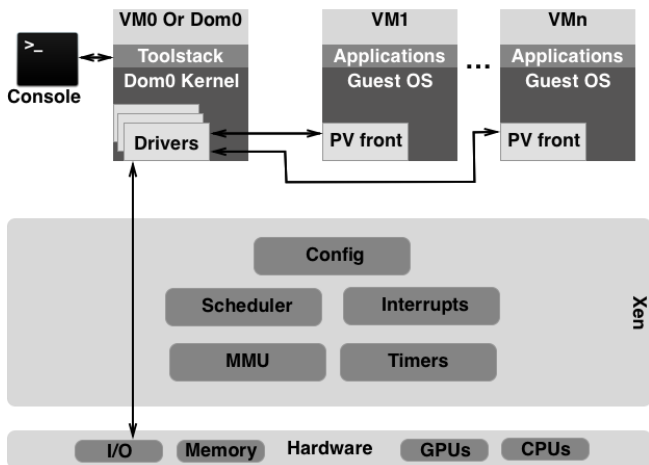

Figure 1. XEN Architecture.

In the case of XEN in Figure 1. , its designers developed a microkernel, placed over the computer's hardware, making possible to run many instances of the operating system. Domain 0 is the privileged VM, containing all the drivers for the hardware and the control platform for the rest of the VMs. As demonstrated in Figure 2. KVM is also a mini kernel, this time completely attached to the Linux kernel, meaning that every distribution after 2.6.20 contains the KVM hypervisor by default. The difference is that instead of using a middleware with drivers, as XEN does, KVM has excellent hardware support.

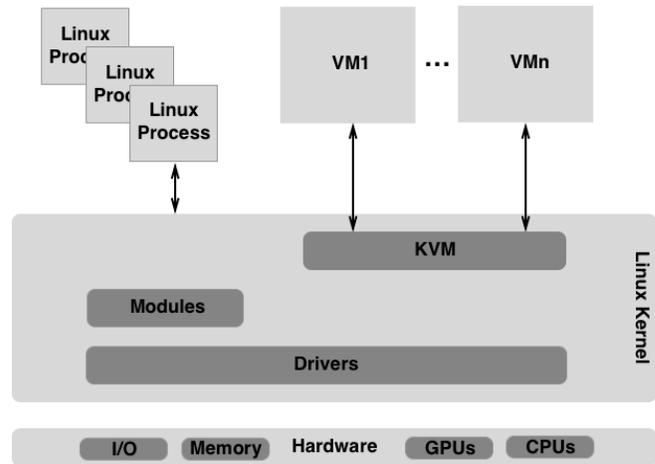

Figure 2. KVM Architecture.

### B. Hypercalls

In either case, as the hypervisor is responsible for monitoring all privileged actions, VMs have to transfer control into the hypervisor to execute sensitive instructions. This procedure is materialized by hypercalls. The latter are very similar to system calls in conventional operating systems. A software interrupt transfers control from the VM into the hypervisor, where every operation is validated and then executed. After the operation is completed, the control returns to the VM that made the call initially. Hypercalls, as system calls, differ depending if the architecture is x86 or x64. Their structure is similar to system calls, including parameter passing (for example a XEN hypercall definition: HYPERVISOR_mmu_update(const struct mmu_update reqs[], unsigned count, unsigned *done_out, unsigned foreigndom)).

### C. Graphical Processing Unit Acceleration

The creation and usage of more computational resources demanding algorithms, along with the birth of big data, pushed the worldwide community towards parallel computing. As CPUs can be too expensive, the scientific community turned to GPUs. Modern GPUs have an architecture that enables them to make fast simple mathematical and logical calculations, using multiple cores, which were commonly used for graphics representation. When a medium ranged GPU can offer more than 1000 cores, it is more energy and cost efficient than any other CPU antagonist. There are two technologies commonly used, NVIDIA's Compute Unified Device Architecture (CUDA) [19] and AMD's High Performance Computing [4], which relies on OpenCL™ cross-platform programming language [45].

It is feasible to access GPUs at high performance, within all the major hypervisors, thus taking advantage of all the benefits that Cloud Computing platforms can offer, along with the accessibility of on-demand accelerator hardware. This procedure is called GPU passthrough technology and permits any virtual machine to access one or more GPUs. It is accomplished using two strategies, either API remoting with device emulation or PCI passthrough. Recently, researchers proved that GPU passthrough technology can take advantage of 96-100% of the base systems performance [39].

## III. RELATED WORK

Several attempts to track, disable or counter the malicious insider threat have been recorded. However, the majority of these solutions achieve their goal by focusing on a very specific aspect of the cloud, such as the employees or the network, while only a minority of them aim to provide a general purpose solution [3][11][15][28][32][33][35][36]. Solutions that propose monitoring of system calls and invocation of statistical methods for identifying normal and malicious acts are [2][8][12][21][23][25][29].

Coull's work [6] has inspired the initial CROW method. They used the system calls as a series of genes and made use of the Smith Waterman algorithm. However, they did not use entire patterns, something that has resulted in many false positives and false negatives. Compute Unified Device Architecture (CUDA) involvement was proposed by Ioannidis et al. [37] for executing Snort [31]. In [1] Haddad et al. propose a scheme aiming to detect network attacks, consisting of Snort for signature based detection and Support Vector Machine (SVM) for anomaly detection. Furthermore, Vasiliadis et al. [38] proved that GPU acceleration is so efficient that can be used by malicious parties in order to increase the robustness of malware against analysis and detection.

Milenkoski et al. [18] created "HInjector", which is a customizable framework, able to inject hypercall attacks during regular operation. This is the reason why Wang et al. [40], created a mechanism that aims to protect the hypercall interface by preventing untrusted hypercalls from running, using randomization techniques.

## IV. THREAT MODEL

According to Maybury et al. [16] the term "insider", for an organization system, applies to anyone with approved access, privilege, or knowledge of the information system and its services and missions. "Malicious insider" is defined as someone motivated to adversely impact an organization's mission through a range of actions that compromise information confidentiality, integrity, and/or availability taking advantage of his/her privileges. This terminology covers mostly traditional IT systems. A modern update would be that a malicious insider is someone who acts either actively or passively. In the first case, an active malicious insider is motivated by himself to harm an organization. A passive malicious insider, is a victim of phishing or other social attack (social engineering, phishing, etc.), whose actions are orchestrated by an external attacker. Consequently, he uses his privileges to harm an organization, without his will.

In the case of Cloud Computing, we define as insider an entity who: (a) Works for the cloud host, (b) Has privileged access to the cloud resources and (c) Uses the cloud services. All cloud insiders are mostly privileged users, who either at will or not, compromise a Cloud infrastructure's security. Depending on their privileges, the impacts from their actions vary from a temporary break of network or a service, to users' privacy violation or loss/exposure of data. There is infrastructure related information, such as the network topology that can be extracted only by privileged users. For example, a malicious user will try to make a map of all available VMs, in order to choose his next target, which will give him more information and will help him to violate the security of a Cloud infrastructure or a user's privacy.

As hypercalls are like system calls, this gives the ability to the potential attackers to perform or inject hypercall attacks, which can take any form known from system calls, such as argument highjacking or mimicry [44]. Another tactic commonly used, is to fake a series of hypercalls with ultimate purpose to sniff the information from other VMs. In addition to that, Cloud infrastructures lack physical isolation by default because of their architecture, something that offers the opportunity to several VMs to get information from shared sources of the Cloud ecosystem such as memory (cache or main memory) retrieving personal information for the co-residents. Ristenpart et al. [26] first proved this concept by performing cross VM side channel attacks on Amazon EC2, measuring in that way the activity of other users. Similarly, Rochsa and Correia [27] proved that, by using the memory of a VM, sensitive information about its users can be acquired, such as social security number, credentials and other personal information.

There are other cases, where attackers combine utilities and tools, whose functionalities are commonly perceived as benign, in order to perform an attack. An example of such a case are the commands "nslookup", "ping" and the nmap tool, which can access publicly available information regarding network topologies and OS, for a specific ecosystem of VMs. The results from those commands orchestrate a "co-residence" or "co-tenancy" attack [26]. Furthermore, following the way of thinking of commonly employed Advanced Persistent Threats, this kind of information may prove useful in the future as it leads to exploits of vulnerabilities relevant to OS version and the other characteristics of a VM. Another kind of attack that can be performed inside a virtual network, is a network stress attack named "smurf" where the attacker launches numerous ping requests, thus congesting the corresponding public and private interfaces and eventually causing Denial Of Service. Modified And Deterring Cloud Realtime Observation Wards

### A. Overview

The proposed scheme, namely MAD CROW is a modified and improved version of another proposed solution [22]. Its goal is to facilitate detection of malicious privileged users in the cloud, regardless of if they use the non-privileged VM, or the privileged VM Domain0 or Dom0 or VM0. It also provides functionality of traditional IDS implementations by individually monitoring the health of each employed VM. Its unique feature is the use of both hypercalls for the non-privileged VMs and system calls for Dom0. To the best of our knowledge, it is the first of this kind. It's high level architecture is depicted in Figure 3.

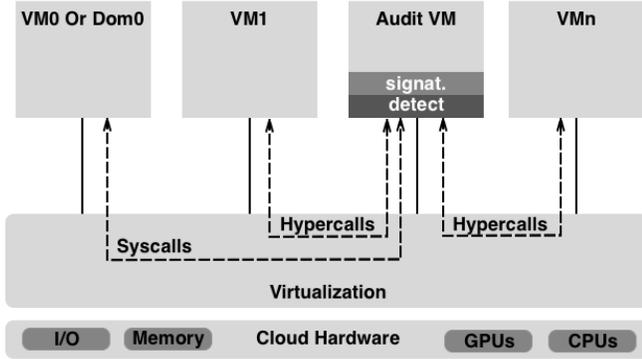

Figure 3. The MAD CROW Architecture.

As highjacking techniques exist and can fool the system call tracing inside a VM, the proposed scheme makes use of a filtering system for hypercalls. Each time a hypercall is initiated, it is recorded in the hypercall sequence of the specific VM that made the call into the audit VM. So, we propose a unified mechanism, which has signatures in terms of hypercall sequences relevant to the operations of each VM. This mechanism constantly detects the hyper calls through the hypervisor, using GPU acceleration instead of CPU usage. Whenever an attack signature is detected by the audit VM, a security alert is generated for the security officers to act. In the case of Dom0 as it is the privileged VM and the highest in the hierarchy, being able to damage the entire cloud infrastructure, the system calls detection is mandatory. To be specific, a mechanism is installed inside the privileged VM and detects its system calls through the kernel. Whenever something abnormal is detected, an alert reaches the audit VM. In both cases, the detection is achieved using GPU acceleration and passthrough technology, in both the Audit VM and privileged VM tracking mechanism.

The sub-system, which implements the audit mechanism, is responsible to monitor the health of each of the VMs either through hypercalls (non-privileged) or through system calls (privileged). Additionally, it generates new attack signatures, based on the hypercall and system call patterns of the attacks. The proposed scheme makes also use of a detection module, which monitors each VM and utilizes the attack signatures for computing their similarity with the sequences of hypercalls generated by the non-privileged VMs. In the case of the privileged VM, the same monitoring is achieved using system calls attack signatures. Calculating the similarity score is a very intense procedure, in computational terms, especially in terms of CPU and RAM.

With respect to GPU passthrough technology, our approach focuses on transferring the majority of the introduced overhead to the GPUs. Consequently, the rest of the computational resources of the infrastructure remain almost idle in terms of usage so as to serve the needs of the other users. This procedure has become possible through the architectures of NVIDIA's Compute Unified Device Architecture (CUDA) and AMD's High Performance Computing, which uses OpenCL™ cross-platform programming language [45]. Both are parallel computing platforms that provide access to the virtual instruction set and memory of GPUs.

B. *Attack Signature Generation*

The attack signature generation process is very similar to the CROW methodology [22], but with one major difference. This time we track system calls, for the privileged VM, and hyper calls for all other VMs. The methodology is very simple and intuitive. A significant number of hypercalls and system call patterns is collected, following multiple executions of the same attack. Then, we make use of the Smith Waterman algorithm [30], to process our data. Each hypercall and system call consists of symbols, drawn from a finite discrete alphabet. So, our goal is to find the longest common subsequence to all sequences in a set of sequences, making the Smith Waterman algorithm an excellent choice for our purpose.

The signature extraction is very similar to malware analysis, since the attack is known a priory. Thus, the malware is executed several times in order to get the corresponding signatures. More specifically, the algorithm runs in pairs of sequences of the hypercalls or system calls for the same attack. Then, the number of sequences is reduced to half, using the best similarity match either for hyper calls or for system calls. After all results have been processed, the attack signature is generated. It must be stressed that the privileged VM is able to execute a significant number of attacks on its own, while all the others can both act alone or even cooperate in order to achieve a successful attack. Consequently, according to Figure 4. the proposed methodology can retrieve the appropriate information and when all the segments of an attack are collected to signal an alarm, even though other benign executions interfere and create noise in the sequences of either hypercalls or system calls. We should not forget that simple commands, such as "nslookup" are harmless on their own, but when combined with others may result in mapping an entire network ecosystem [26].

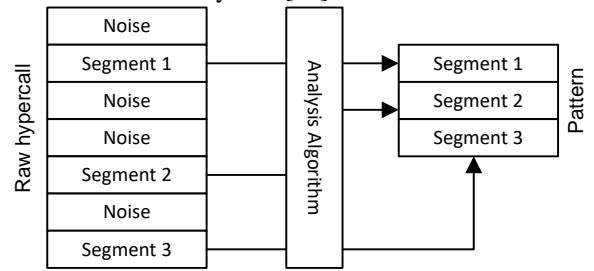

Figure 4. The segments of the attack pattern are found through the hypercall sequence using as analysis the Smith-Waterman algorithm

## C. Detection

The attack signatures created from the former procedure, either as sequences of hypercalls or as sequences of system calls, are used for the detection of potential malicious acts. Specifically, the audit VM, keeps signatures in a database. To achieve the detection of an attack against the VM or the cloud infrastructure itself, the hypercalls of the VMs and the system calls of the privileged VM are monitored and forwarded to the detection module.

Its task is to identify the attack segments into the entire sequence of hypercalls or system calls, avoiding the possible noise that has been created by various other irrelevant system procedures and thus making the same steps as the attack signature generation. In the case where all the segments of an attack are identified, then an alert in the audit VM is triggered. This alert motivates the operators of the audit station to take immediate action and enforce the employed policy.

It must be noted that even in cases where the attack segments are executed in different VMs, which is a typical choice of attackers in order to avoid detection, the proposed scheme will again detect the attack. Additionally, a handshake, between the audit station and each of the VMs, is initiated every two seconds in order to update the audit station about VM communication and thus protect the system from potential actions that aim to hide an attack.

## V. CONCLUSIONS AND FUTURE WORK

Considering modern IDS systems do not focus on cloud insider attacks, the MAD CROW detection method has been proposed. It utilizes both hypercalls and system calls to detect privileged user attacks. The detection mechanism is based on Smith Waterman algorithm, adapted in a parallel implementation, usable by any GPU architecture and passthrough technology.

Currently, we are experimenting with different implementations and GPU setups, willing to achieve maximum stability, efficiency and productivity. Our experimentation includes different machine learning techniques and feature extraction that would allow us to improve the signature generation mechanism and consequently the accuracy of our detector.

## ACKNOWLEDGMENT

This work has been partially supported by the Research Center of the University of Piraeus.